\documentclass[10pt]{article}
\usepackage{amsmath,bbold,graphicx,array,slashed,amssymb,multirow,cite}
\usepackage[toc,page]{appendix}
\usepackage[normalem]{ulem}
\usepackage{mathrsfs}
\usepackage{pstricks}
\usepackage{color}
\usepackage{simplewick}
\newcommand{\tr}{\text{Tr}}
\newcommand{\Tr}{\text{Tr}}

\newcommand{\ud}{\mathrm{d}}
\newcommand{\Wil}{\mathcal{W}}
\newcommand{\matM}{\mathbb{M}}
\newcommand{\matK}{\mathbb{K}}

\title{Equivalence of Meson Scattering Amplitudes in Strong Coupling Lattice and Flat Space String Theory}
\author{Adi Armoni, Edwin Ireson, Davide Vadacchino\\ \\
	\textit{Department of Physics, Swansea University,}\\ \textit{Swansea SA28PP, United Kingdom}}
\begin{document}
	\maketitle

\begin{abstract}
  We consider meson scattering in the framework of the lattice strong coupling expansion. In particular we derive an expression for the 4-point function of meson operators in the planar limit of scalar Chromodynamics. Interestingly, in the naive continuum limit the expression coincides with an independently known result, that of the worldline formalism. Moreover, it was argued by Makeenko and Olesen that (assuming confinement) the resulting scattering amplitude in momentum space is the celebrated expression proposed by Veneziano several decades ago. This motivates us to also use holography in order to argue that the continuum expression for the scattering amplitude is related to the result obtained from flat space string theory. Our results hint that at strong coupling and large-$N_c$ the naive continuum limit of the lattice formalism can be related to a flat space string theory.
\end{abstract}
\clearpage
\tableofcontents
\clearpage
\section{Introduction}

Studying the strong coupling regime of QCD is a hard, long-standing problem for which we are keenly awaiting a solution. There are several approaches to tackle this problem, among them the lattice strong coupling expansion and string theory.

The lattice strong coupling expansion, formally analogous to a high temperature expansion in statistical mechanics, is a technique that allows us to explore the non-perturbative regime of a lattice field theory with an expansion in powers of the inverse coupling. In this framework, it is possible to show that strongly coupled lattice regularized YM theories exhibit both a mass gap and a non vanishing string tension. Despite the impossibility of relating the latter to a continuum theory, they are inspiring and qualitatively very important. The gauge/gravity correspondence, another important approach, suffers from various problems too: it is difficult to extend its results beyond large-$N_c$ and strong coupling. More importantly, it is not known what the gravity dual of QCD is, if any. However, the two approaches show some superficial similarity, and it is an intriguing prospect to investigate just how closely the two methods actually are.

In this manuscript we will argue that in the large-$N_c$ limit and at strong coupling, certain scattering problems in lattice field theory and string theory share strikingly similar methods of computation, and subsequently the same result, in this corner of parameter space. More precisely, we consider $SU(N_c)$ scalar Chromodynamics with  $N_f$ scalars transforming in the fundamental representation in the 't Hooft limit. We argue that the $n$-point function of meson operators of the form
\begin{equation}
\langle \phi_{a_1} ^\dagger \phi_{a_1}^{\phantom{\dagger}} (x_1)  \phi_{a_2} ^\dagger \phi_{a_2}^{\phantom{\dagger}} (x_2)  ...  \phi_{a_n} ^\dagger \phi_{a_n}^{\phantom{\dagger}} (x_n) \rangle \label{n-point}
\end{equation}
yield in the naive continuum limit of the lattice the same result as flat space string theory. In momentum space the $4$-point function is given by the celebrated Veneziano amplitude.

Our main result is that the above $n$-point function \eqref{n-point} is given on lattice by 
\begin{equation}
   \sum_{l}\frac{\kappa^l}{l}\sum_{ x_1\dots x_n \in C_l} \exp (-\sigma {\cal A}) \,,
\end{equation}
and becomes in the continuum
\begin{equation}
\int_0^\infty \frac{dT}{T} \int\limits_{x(0)=x(T)}^{ x_1\dots x_n \in C} [Dx]\,\exp\left( -\int dt\, \frac{\dot{x}^2}{4} \right) \exp (-\sigma {\cal A}) \, ,
		\label{continuum_limit}
\end{equation}
where the sum is over all closed contours $x(t)$ of length $T$ that include the points $x_1,...,x_n$ and ${\cal A}$ represents the minimal area enclosed by the contour ${\cal C}$. In addition, noting that the strong coupling regime induces a large value for the string tension, we will show that the same expression is computed by flat space string theory
\begin{equation}
 \int_{ x_1\dots x_n \in \partial X} DX \exp\left( -\sigma{\int d^2\sigma\,  \partial_\alpha X_\mu \partial ^\alpha X^\mu } \right ) \, ,
\end{equation}
where $X^\mu$ are string worldsheets with the topology of a disk. The points $x_1, x_2, ... , x_n$ lie on the boundary of the disk. All of these expressions are then expected to produce the famous Veneziano amplitude as the end point of their computation, as a result.

Our key observation is closely related to a statement by Makeenko and Olesen \cite{Makeenko:2009rf} that if in a gauge theory all Wilson loops (however small) admit an area law, scaling with a large string tension, then the meson scattering amplitude is given by the Veneziano amplitude. Since at strong coupling the lattice yields an area law for all Wilson loops, we anticipate a close relation with flat space string theory. 

The paper is organised as follows: in section 2 we derive the expression for meson scattering on the lattice and discuss its continuum limit. In section 3 we re-derive the obtained expression in the continuum. Furthermore, we argue, relying on the result by Makeenko and Olesen  \cite{Makeenko:2009rf} and purely within field theory, that meson scattering for a theory with area behaved Wilson loops is given by the Veneziano amplitude. In section 4 we use the gauge gravity correspondence to relate the continuum limit expression for meson scattering to flat space string theory. In section 5 we discuss our results. Finally, we have included two appendices to discuss some finer points not previously explained, on reparametrisation invariance and on fixed boundaries in string theory.

\section{On the Lattice: the Hopping Expansion}
\subsection{Setting up the Hopping Expansion}
In $4$ Euclidean space-time dimensions, the continuum Lagrangian density of a complex scalar field coupled to a $SU(N_c)$ Yang-Mills field $A_\mu$ is 
\begin{equation}\label{eq:euclidean_action}
	S_E[\phi^\dag,\phi,A_\mu] = S_{\phi} + S_{\rm YM} = \int \ud^4 x \left[ (D_\mu \phi)^\dag (D^\mu \phi) + \frac{m^2}{2} \phi^\dag \phi  + \frac{1}{4 g^2} \Tr( F_{\mu\nu} F^{\mu\nu}) \right]
\end{equation}
where $\mu=0,\cdots,3$ labels the direction in space-time, $\phi$ lies in the fundamental representation of $SU(N_c)$, and $m$ is its bare mass. The covariant derivative $D_\mu = \partial_\mu + \imath A_\mu$ implements the minimal coupling between $\phi$ and the gauge field $A_\mu$. The dynamics of the latter is dictated by the last term where $F_{\mu\nu}$ is the field strength tensor, $g$ the gauge field coupling constant and the trace is over colour space. Since the following discussion can be generalized to the case of $N_f$ flavours, we limit to $N_f=1$ for simplicity.

We now discretise the action on a hypercubic space-time lattice $\Lambda = a^d \mathbb{Z}$ with sites $n$, links $(n,\hat{\mu})$ and plaquettes $(n,\hat{\mu},\hat{\nu})$. For the gauge part, we adopt the Wilson action
\begin{equation}
	S_{\text{\tiny\rm YM}} = \beta \sum_{n\in\mathbb{Z}^4}\sum_{\mu >\nu}\left[ \mathbb{1} - \frac{1}{N_c} Re \Tr ~U_{\mu\nu}(n) \right]
\end{equation}
where $\beta=2N_c/g^2$ and 
\begin{equation}
	U_{\mu\nu}(n) = U_\mu(n) U_\nu(n+\hat{\mu}) U^\dag_\mu(n+\hat{\mu}+\hat{\nu}) U^\dag_\nu(n+\hat{\nu})
\end{equation}
is the oriented product of lattice parallel transporters $U_\mu(n)$ along links around an elementary plaquette. 

For $S_\phi$, with a naive discretisation for the derivative
\begin{equation}
	\partial_\mu \phi \to \frac{\phi(n+\hat{\mu} a) - \phi(n)}{a}
\end{equation}
and using the freedom to redefine the field as follows $\phi\to\phi\sqrt{\kappa}/a$, one obtains
\begin{equation}
	S_\phi = \sum_{n,l\in\mathbb{Z}^4} \sum_{b,c=1}^{N_c} \phi^\dag_{n,b} ~ \matK_{nb,lc}\left[ U \right]~ \phi_{l,c}
\end{equation}
where  
\begin{equation*}\label{eq:hoppingpar}
	\kappa = \frac{1}{(2d + \hat{m}^2)}
\end{equation*}
is the \emph{hopping parameter} and $\hat{m}= ma$. The \emph{hopping matrix}
\begin{equation}\label{eq:hopmatrix}
\matK_{bn,cl}\left[U \right] = \delta_{b,c}\delta_{n,l} - \kappa~\sum_{\hat{\mu}} U_\mu(n)_{bc} \delta_{l,n+a\hat{\mu}} = \left( \mathbb{1} - \kappa \matM[U] \right)_{bn,cl}~,
\end{equation}
has indices in colour space and space-time and only depends on the discretised gauge field $U$. Note that, from its definition, the hopping matric has nonvanishing elements only for lattice neighbours.

The classes of possible gauge invariant objects present in this theory are traced ordered products of link variables, 
\begin{equation}
	\Wil(\mathcal{C}) = \Tr \left( \prod_{l\in\mathcal{C}} U_l \right)
\end{equation}
also known as Wilson loops, correlators of parallel transported fields $\phi(n_1)^\dag U\cdots U\phi(n_2)$, and local invariant products of $\phi$ fields like $\phi^\dag \phi(n)$. Contact with alternative definitions of Wilson loop, namely that with a factor of $1/N_c$ in the front of the trace, can be made by rescaling the scalar field with $\sqrt{N_c}$. All the non gauge invariant correlation functions can be shown to vanish on average. In this computation we are interested in connected correlators of $\phi^\dag\phi(n)$ that we interpret as mesons built out of scalar quarks. For the usual reasons of factorisation processes in amplitudes, we believe that this is enough to capture broad features of the processes, and that it could be generalised to more realistic, spinor quarks with some effort.

As usual in QFT and Statistical Mechanics, we can express connected correlation function of local operators as functional derivatives of a generating functional $Z[J,J^\dag]$ with respect to the sources $J^\dag$ and $J$ that lie in the fundamental representation of $SU(N_c)$ at each space-time lattice node. Straightforward algebraic manipulations show that  
\begin{equation}
	Z[J,J^\dag] = \int \mathcal{D}U ~e^{-S_{\rm YM}} ~\det \matK^{-1} \exp\left( \sum_{n_1, n_2} \sum_{a_1, a_2} J^\dag_{n_1, a_1} \matK[U]^{-1}_{n_1 a_1, n_2 a_2} J_{n_2, a_2} \right)~.
\end{equation}

Connected correlation functions of the local operator $\phi^\dag \phi$ are obtained from the following general expression
\begin{equation}\label{eq:corr_con}
	\langle \phi^\dag \phi(n_1) \cdots \phi^\dag \phi(n_k) \rangle_c = \left.\frac{\delta^{2k} }{\delta J^{\dag}_{n_1} \delta J_{n_1} \cdots \delta J^\dag_{n_k}\delta J_{n_k}}\ln Z[J,J^\dag]\right|_{J_n=J^\dag_n=0}
\end{equation}
where the colour indices that have been on the currents in the functional derivative operators are actually summed over at each separate lattice site.

Then the 4-point function is obtained as,
\begin{equation}
	\langle \phi^\dag \phi(n_1) \cdots \phi^\dag \phi(n_4) \rangle_c 
	=\frac{1}{Z[0,0]} \int \mathcal{D}U e^{-S_{\rm YM}}~ \det \matK^{-1}~ \left(\Tr~\matK^{-1}_{n_1,n_2}\matK^{-1}_{n_2,n_3}\matK^{-1}_{n_3,n_4}\matK^{-1}_{n_4,n_1}+ \text{perms.}\right)
\end{equation}
where the simple result
\begin{equation}
\left. \frac{\delta}{\delta J_{n_1,a_1}}\frac{\delta }{\delta J^\dag_{n_2, a_2}} \exp( J^\dag \matK^{-1} J ) \right|_{J^\dag = J = 0} = \matK^{-1} _{n_1a_1, n_2 a_2}[U] 
\end{equation}
has been used and the trace is over color indices. The other terms, denoted collectively by "perms.", are all other ways to Wick contract the fields in the amplitude. Each of these addends can be put in one-to-one correspondence with the permutations of $(n_1,\dots,n_4)$, we now focus on just one of these contributions, and later show how the remaining ones will be taken into account

From the definition in eq.~(\ref{eq:hopmatrix}) the operator $\matK^{-1}[U]$ can be expanded in a geometric series for $||\kappa \matM||<1$, condition that can be shown to be equivalent to $\hat{m}>0$, and we obtain
\begin{equation}\label{eq:inverseprop}
	\matK_{n_1a_1,n_2a_2}^{-1} = (\mathbb{1} - \kappa \matM[U])_{n_1a_1,n_2a_2}^{-1} = \sum_{\mathcal{C}_{n_1,n_2}} \kappa^l (\matM[U]^l)_{n_1a_1, n_2a_2}
\end{equation}
As noted before, the hopping matrix only connects neighbouring sites, and the sum over $\mathcal{C}_{n_1,n_2}$ means a sum over all paths that connect $n_1$ and $n_2$, with $l$ their length. For clarity, we can explicitly write down the first few terms,
\begin{multline}
	\matK_{n_1 a_1,n_2 a_2}^{-1} = \kappa ~\matM[U]_{n_1a_1,n_2a_2} \delta_{n_1,n_2+a\hat{\mu}} + \\ + \kappa^2 ~{\sum}_{m_1,m_2} \delta_{m_1,n_1+a\hat{\nu}} \delta_{m_2,n_2+a\hat{\mu}} \delta_{m_1,m_2} \left( \matM[U]_{n_1 m_1} \matM[U]_{m_2 n_2}\right)_{a_1 a_2} + O(\kappa^3)
\end{multline}
The spatial $\delta$-function in the second term above means that there is a O($\kappa^2$) contribution only if there exists a path connecting $n_1$ and $n_2$ whose length is equal to two elementary links. The intermediary colour indices were omitted in that term, but it is understood that matrix multiplication occurs in colour space also. Therefore, we have implicitly summed over intermediate colour indices and the product of $\matM$ matrices must be understood as a matrix product in colour space and space-time. Generalizing the above reasoning, the propagator between $n_1$ and $n_2$ is thus the sum of the contribution of every path that connects them, each weighted by a power $l$ of the hopping parameter equal to the length of the path.


The above formula tells us that to compute the latter contribution, we have to connect each pair of points with a path: ignoring colour indices, we have
\begin{equation}
	\Tr~\matK^{-1}_{n_1,n_2}\matK^{-1}_{n_2,n_3}\matK^{-1}_{n_3,n_4}\matK^{-1}_{n_4,n_1} = \sum_{l_1,l_2,l_3,l_4} k^{l_1+l_2+l_3+l_4}\Tr~ \matM[U]_{n_1,n_2}^{l_1} \matM[U]_{n_2,n_3}^{l_2} \matM[U]_{n_3,n_4}^{l_3} \matM[U]_{n_4,n_1}^{l_4} 
\end{equation}
where $l_i$ is now the length of the path going from $n_i$ to $n_{i+1}$. Now in this expression the sum is separately over the the four pieces going from $n_i$ to $n_{i+1}$. By restating this sum as a sum over all closed loops starting from either one of the $n_i$ and passing through the remaining three, it includes the permutation terms that were previously neglected. To make contact with later computations, we would like the sum to not involve choosing a definite starting point along the closed paths at hand. This is done so as to sum over distinct geometrical curves: we are currently overcounting each closed path on the lattice by as many times as there are ways to choose a starting point for it, that is, each contributes $l=l_1+\dots l_4$ times. We divide through by this factor and re-express the result as the following, elegantly geometric sum, the form of which is closely related to expressions in the continuum:
\begin{equation}
	\Tr~\matK^{-1}_{n_1,n_2}\matK^{-1}_{n_2,n_3}\matK^{-1}_{n_3,n_4}\matK^{-1}_{n_4,n_1} = \sum_{l}\frac{\kappa^l}{l}\sum_{n_i\in C_l} \Tr~ \matM[U]^l
\end{equation}

From eq.~(\ref{eq:hopmatrix}) and our convention on implicit summation of colour indices, it is easy to show that for closed loops, the product of $\matM[U]$ matrices is an ordered and traced product of link variables, known as Wilson loop\
\begin{equation}
	\Tr~\matM[U]^l =  \Tr\left(~ \prod_{l\in\mathcal{C}} U_l~\right)= \Wil(C_l) ~.
\end{equation}
Therefore
\begin{equation}
	\langle \phi^\dag \phi(n_1) \cdots \phi^\dag \phi(n_4) \rangle_c = \sum_{l}\frac{\kappa^l}{l}\sum_{n_i\in C_l} \langle (\det \matK^{-1})~ \Wil(C_l)\rangle_{\rm YM}
\end{equation}

This result is exact for any value of $N_c$ and of the coupling $\beta$. Now in the limit of large $N_c$ with $\lambda=N_c g^2$ constant, the simplification $\det \matK = 1 + O(1/N_c)$ occurs and the determinant disappears from the above formula, at leading order. The net effect is thus analogous to what one would get in the quenched approximation. 

The vacuum expectation value $\langle ~\Wil(C_l)~ \rangle_\text{\tiny YM}$ can be evaluated in the strong coupling approximation\cite{Munster:1980ab}, $\lambda\sim\infty$ both at finite and infinite $N_c$. In this framework, many observables can be systematically obtained as (convergent) power series of the inverse coupling. To each term of the expansion is associated a specific lattice graph, the latter depending on the observable one wishes to compute. For the Wilson loop, these are the surfaces bordered by the path $\mathcal{C}_l$, and each contribution is weighted by $\lambda^{-\mathcal{A}}$, where $\mathcal{A}$ is area of the surface. The leading order contribution naturally corresponds to the minimal surface, $\langle ~\Wil(C_L)~\rangle_{\text{YM}} \sim \lambda^{-\mathcal{A}}$, with corrections corresponding to fluctuations around it. 

By keeping only the leading terms in the large-$N_c$ and strong coupling approximations, we obtain
\begin{equation}\label{eq:largeNstrong}
	\langle \phi^\dag \phi(n_1) \cdots \phi^\dag \phi(n_4) \rangle_c \simeq \sum_{l}\frac{\kappa^l}{l}\sum_{n_i\in C_l} e^{-\sigma \mathcal{A}},
\end{equation}
where $\sigma = \ln\lambda$ is the string tension at the lowest order in the strong coupling approximation. The above formula is stated for $n_1,\dots,n_4$, but can be trivially generalized to a general number of operators in the correlator.  

The continuum limit of this theory is obtained by sending $a\to0$ with the couplings $(g,\kappa)$ of the theory tuned to their critical values, i.e. for which the system has a second order phase transition. In our case, this happens when $g=g_c=0$ and $\kappa=\kappa_c=1/8$. While there is no difficulty in tuning $\kappa$ to its critical value, for $g$ we would have to cross the so-called roughening transition\cite{Munster:1980ab}, that separates the weak and strong coupling regimes of the theory. In the following, we will then fix $\sigma$ to some non-vanishing value and perform a naive continuum limit, that consists in sending the lattice spacing to $0$ without modifying the coupling. The problem of performing the above sum is then mapped onto a problem of lattice geometry, specifically of random closed loops. We now want to introduce physical units to all the quantities at hand, so we make factors of the lattice spacing $a$ explicit where needed e.g. to measure the mass. We also need to introduce factors of the lattice simulation time $\tau$, measuring the length of paths $l$, which \textit{a priori} are different quantities. By noting that
\begin{equation}
	\kappa^l = \left(\frac{1}{2d}\right)^l \left( \frac{1}{1 + \frac{m^2 a^2}{2d}}\right)^l \simeq \left(\frac{1}{2d}\right)^l e^{-l\frac{m^2 a^2}{2d}}
\end{equation}
for $\hat{m}=m a\ll\sqrt{2d}$, eq.~(\ref{eq:largeNstrong}) can be recast as,
\begin{equation}
	\langle \phi^\dag \phi(n_1) \cdots \phi^\dag \phi(n_4) \rangle_c = \sum_{l=0}^\infty \frac{\tau}{l\tau} e^{-\frac{l\tau}{2d\tau}m^2 a^2} \sum_{n_i\in C_l} \left(\frac{1}{2d}\right)^l e^{-\sigma \mathcal{A}}
	\label{eq:latticeamplitude}
\end{equation}
In this expression we have conspicuously written terms depending on the quantities $l\tau$ and $m a$. Expressed this way, we can now observe the continuum limit of our amplitude.

\subsection{Continuum limit of theories of random paths}

The study of random walks on discrete spaces is a canonical problem in probability theory, and their limiting behaviour to the continuum well-understood. They give rise to Brownian motion problems, typical of diffusive behaviour in physical systems. We will use ideas found in a review of the subject \cite{Itzykson:1989sx} to evaluate the continuum limit of the computation above. Crucially it is explained that, in order for the continuum limit to be non-trivial, we require the lattice spacing $a$ and the simulation time-step $\tau$ to obey
\begin{equation}
\tau = \frac{1}{2d} a^{2}
\end{equation}
and vanish simultaneously according to this relation.

Substituting this relation in the expression above, we obtain:
\begin{equation}
\langle \phi^\dag \phi(n_1) \cdots \phi^\dag \phi(n_4) \rangle_c = \sum_{l=0}^\infty \tau \frac{1}{\tau l} e^{-\tau l {m^2}} \sum_{n_i\in C_l} \left(\frac{1}{2d}\right)^l e^{-\sigma \mathcal{A}}
\end{equation}

The continuum limit acts on two different sums, one controlled by the lengths of the contours we are looking at (a very straightforward one-dimensional sum) and another over all shapes of contours of a given length (a bigger and more complicated space).

The first sum converts to an integral very automatically by loosely applying Riemann's definition of the integral: for a function $f$ which we assume well behaved enough,
\begin{equation}
\sum_{l=0}^\infty \tau f\left( l\tau\right) \underset{\tau\rightarrow 0}{ \longrightarrow }\int_0^\infty dT f(T)
\end{equation}

The second sum converts to a \textit{path integral}, in a statistical sense. For any well behaved operator,
\begin{equation}
\sum\limits_{n_i\in C_l} \left(\frac{1}{2d} \right)^l  O(C_l) \longrightarrow \int_{x_i\in C_T} [Dx]\,\, \exp\left( -\int_0^T dt \,\,\frac{1}{4}\left( \frac{dx}{dt}\right)^2\right) O(C_T)
\end{equation}
where the path integration measure $[Dx]$ is weighted such that we sum over all possible geometrically distinct closed curves of length $T$, all of which pass through the marked points $x_i$ defined by the operator insertions. The justification for this is found in the aforementioned review \cite{Itzykson:1989sx}.

Combining these notions we find that

\begin{equation}
\langle \phi^\dag \phi(x_1) \cdots \phi^\dag \phi(x_4) \rangle_c \longrightarrow \int_0^\infty \frac{dT}{T}  \int_{x_i\in C_T} [Dx] e^{ -\int_0^T dt \left( \frac{\dot{x}^2}{4}+m^2\right)}  e^{-\sigma \mathcal{A}[\mathcal{C_T}]}
\end{equation}

The end product is an expression that does not seem to be reparametrisation invariant, but we have glossed over a subtlety concerning such matters, the details of which are summarised in Appendix \ref{appA}. The remarkable thing to notice here is that this expression is readily available to derive in the continuum from an altogether not entirely unrelated framework, the Worldline formalism. 

\section{In the Continuum: Worldline Path Integrals, Area Functionals}

\subsection{Worldline formulation of the amplitude}

The worldline formalism can be thought of much in the same way as the lattice hopping expansion: functional integrals can be expressed as an averaging process over all configurations of a field, or in a more geometric way as an average over all trajectories that the particle state of this field can take as it moves through space. Let us see how this is done in the continuum.

We want to compute the following object

\begin{equation}
	\left<\prod\limits_{i=1}^n  \phi(x_i)^\dagger\phi(x_i)\right> =\frac{1}{\mathcal{Z}} \int DAD\phi D\phi^\dagger\,\, e^{-S_{YM}}   \prod\limits_{i=1}^n \phi(x_i)^\dagger_{a_i} \phi(x_i)_{a_i}e^{-\int d^dx \left( (D\phi)^\dagger (D\phi)+m^2\phi^\dagger\phi\right) }
\end{equation}

By introducing a current in the scalar action, $J\phi^\dagger \phi$, this expectation value can be represented by
\begin{align}
\left<\prod\limits_{i=1}^n  \phi(x_i)^\dagger_{a_i}\phi(x_i)_{a_i}\right> &=\prod\limits_{i=1}^n \frac{\delta}{\delta J(x_i)}{\bigg\rvert}_{J=0}\frac{1}{\mathcal{Z}} \int DAD\phi D\phi^\dagger\,\, e^{-S_{YM}}e^{-\int d^dx\left(  (D\phi)^\dagger (D\phi)+(m^2+J)\phi^\dagger\phi\right) }\nonumber\\
&= \prod\limits_{i=1}^n \frac{\delta}{\delta J(x_i)}{\bigg\rvert}_{J=0}\frac{1}{\mathcal{Z}} \int DA \det\left( D^\dagger D  + m^2 + J \right)^{-N_f}\\
&=\prod\limits_{i=1}^n \frac{\delta}{\delta J(x_i)}{\bigg\rvert}_{J=0}\frac{1}{\mathcal{Z}} \int DA \exp\left( -N_f \Gamma[A] \right) \nonumber
\end{align}
where $\Gamma[A]$ is the effective action for the scalar, $\log \det\left( D^\dagger D + m^2 + J \right)$. We can rewrite $\Gamma$ as a worldline integral: by using $\log \det A = \tr \log A$ and using an integral representation for the logarithm of an operator,

\begin{equation}
	-\log\det\left( D^\dagger D+ m^2 + J \right) = \Tr\int_0^\infty \frac{dT}{T} \exp\left(-\vphantom{\int}T\left(  (\partial+i A)^\dagger(\partial+i A) + m^2 + J\right) \right)
\end{equation}

The advantage of the formula above is that it involves Gaussian functions of operators, which makes it easy to take its trace. For instance, in momentum space:

\begin{equation}
-\log\det\left(D^\dagger D + m^2 + J \right) = \Tr\int_0^\infty \frac{dT}{T} \int \frac{d^d p}{(2\pi)^d} \exp\left( \vphantom{\int} -T\left( (p + A)^2 + m^2 + J\right) \right) 
\end{equation}

This can then be Fourier-transformed into position space. It was shown in \cite{Makeenko:2002uj} that this is achieved by viewing $T$ as the proper length of a particle propagating in a closed loop. One needs to sum over all shapes of this loop to take the full trace of the operator, i.e. we perform a functional Fourier transform with the kernel $\int\limits_{x(0)=x(T)} [Dx] \exp\left( {i \int_0^T dt\,\, \dot{x}(t)\cdot p}\right) $. The exact meaning of the functional integration measure $[Dx]$, which is purported to sum over geometrically distinct embeddings in a reparametrisation invariant way, is a technicality we cover in Appendix \ref{appA}.

The result is that this transformation produces the following worldline representation of the determinant:

\begin{equation}
-\log\det\left( D^\dagger D + m^2 + J \right) = \Tr\int_0^\infty \frac{dT}{T} \int_{C_T} [Dx]\, e^{i\oint dx\cdot A(x)}  \exp\left(-\int_0^T dt \, \left( \frac{\dot{x}^2}{4} + m^2 + J(x(t))\right) \right) 
\end{equation}
This expression already seems familiar- we are averaging a Wilson operator over all shapes and sizes of contours, weighted by a certain exponential weight.

The weight, the integral appearing in the second exponential, is known to be (at $J=0$) an expression for the proper length of the trajectory of a massive particle. It is derived from the usual action by the exact same process that the Polyakov string action is derived from the Nambu-Goto action in string theory. Appendix \ref{appA} contains more information on that subject.

We have therefore expressed again our functional expressions in terms of Wilson operators. As before we use the hypothesis of large $N_c$ to justify simplifying the full amplitude, here by taking $e^{-N_f\Gamma} \simeq 1-N_f\Gamma$ \cite{Armoni:2008jy}. The zeroth order term does not depend on $J$ so its contribution vanishes, leaving only the first term, $\Gamma$, which we then average over all gauge configurations. In total:

\begin{equation}
	\left<\prod\limits_{i=1}^n  \phi(x_i)^\dagger_{a_i}\phi(x_i)_{a_i}\right> \propto\prod\limits_{i=1}^n \frac{\delta}{\delta J(x_i)}{\bigg\rvert}_{J=0}\int_0^\infty \frac{dT}{T} \int_{C_T} [Dx]\, \left<\mathcal{W}(C) \right>  e^{-\int_0^T dt \, \left( \frac{\dot{x}^2}{4} + m^2 + J(x(t))\right) }\label{worldline}
\end{equation}

We can now take the functional derivatives resulting from the operator insertions. Schematically:
\begin{equation}
	\frac{\delta}{\delta J(x_i)}{\bigg\rvert}_{J=0} e^{-\int_0^T dt \,J(x(t))}=\int_0^T dt_i\,\, \delta(x(t_i)-x_i)
\end{equation}
and this expression, when inserted in the path integral $\int_{C_T} [Dx]$, forces that the closed contours $C_T$, in addition to being of proper length $T$, should also pass through point $x_i$ for some value of the parameter. The operator insertions therefore act as ``pins" for the contours we are summing over, much like on the lattice.

Finally, in order to truly find the analogue of Eq.(\ref{eq:latticeamplitude}), we need to introduce a similar assumption to that of the strong-coupling leading order expansion on the lattice. In effect, we want to enforce that every Wilson loop is area-behaved when their expectation value is taken, which requires computing the minimal area of a surface bound by the contours we are summing over. On the lattice, this is a finite problem, but in the continuum we encounter some difficulties, which we will come to address later. Symbolically, let us assume for now that we can perform the following identification: for every loop size $T$,
\begin{equation}
\left\langle \exp\left( i\oint_{\mathcal{C}_T} A\right)  \cdot dx \right\rangle \longrightarrow \exp\left( -\sigma \mathcal{A}[\mathcal{C}_T]\right) 
\end{equation}
and write the final form of the worldline amplitude, with $\sigma$ some effective tension parameter presumably controlled by the gauge coupling
\begin{equation}
	\left<\prod\limits_{i=1}^n  \phi(x_i)^\dagger_{a_i}\phi(x_i)_{a_i}\right> \propto\ \int_0^\infty \frac{dT}{T} \int\limits_{x_i\in C_T} [Dx]\,   e^{-\int_0^T dt \, \left( \frac{\dot{x}^2}{4} + m^2\right) } e^{-\sigma\mathcal{A}[\mathcal{C}_T] }
\end{equation}
This is precisely the result derived previously from the lattice, a geometrical formulation for the amplitude at hand.

To make this rigorous we need to explain precisely how one computes this minimal area expression.



\subsection{Area Path Integrals and the Worldline: the Douglas Functional}

Given a curve in space, we are tasked to find an analytic way of expressing the area of the surface, bounded by the curve, with the least area. This is called the problem of Plateau, and the first constructive demonstration of a solution was given by Douglas \cite{Douglas} with his now-famous variational set-up. Indeed, this striking demonstration earned him the very first Fields medal. It then appeared that Douglas-like functionals were a direct result of performing open string theory calculations with fixed boundaries, which is altogether not entirely surprising: classical string theory is precisely a variational problem of extremal area. Let us see this in more detail and observe how it applies to our computation:

Consider the following actions, dependent on a closed curve $x(\tau)$, whose worldline is mapped conformally to the real line. We write the variational problem with path integration, over all possible reparametrisations $\theta$ of this embedding:
\begin{align}
	Z[x]&=\int D\theta \exp\left( -S_{\text{Doug.}}[\theta,x]\right) \\
	&=\int D\theta \exp\left( \frac{1}{2\pi}\iint_{-\infty}^{\infty} d\tau d\tau'\dot{ x}^\mu (\tau) \log\left(\theta(\tau)-\theta(\tau')) \right)  \dot{ x}_\mu (\tau')\right) \label{douglas} \\
	&=\int D\theta \exp\left( -\frac{1}{4\pi}\iint_{-\infty}^{\infty} d\tau d\tau' \frac{\left( x^\mu (\theta(\tau)) - x^\mu (\theta(\tau'))\right) ^2}{\left( \tau-\tau'\right) ^2}\right) 
\end{align}
They are equal trivially by integration by parts and relabelling of the integration variables. The claim by Douglas is that, when minimised with respect to the parameter $\theta$, an arbitrary reparametrisation of the boundary, the saddle point configuration will produce the minimal area bounded by the geometric curve defined by the mapping $x$. It is difficult to prove so \textit{ab initio} but some sense of the formula can be made by deriving it from string theory. The fact it can be derived from string theory should not be too surprising- the string action is precisely a variational problem of minimal areas, but it has \textit{a priori} unfixed spacetime boundaries. Performing string path integration with a fixed boundary, it is known \cite{Fradkin:1985qd}, produces amongst other things the Douglas functional. We have transcribed a derivation of this fact in Appendix \ref{appB}.

We use it here as an explicit, analytical formula for the sought-after area that our strong coupling expansion requires. In doing the following,
\begin{equation}
	\mathcal{A[\mathcal{C}]}=\text{SP}\,S_{\text{Doug.}}[x] \, ,
\end{equation}
namely, the area is the saddle point of the Douglas action. The authors of \cite{Makeenko:2009rf} were able to perform further path integration and derive that the meson n-point function they were computing was proportional to a Koba-Nielsen integral. Let us see how this is done.

\subsection{Koba-Nielsen integrands using the Douglas action}

Surprisingly, the inclusion of this area action in the worldline formalism for the meson amplitude, produces expressions very similar to Koba-Nielsen integrals (at least for on-shell momenta), and so in the case of four mesons we obtain the celebrated Veneziano amplitude, without \textit{a priori} doing any proper string theory.

First, we need to move to momentum space amplitudes, which is done in a decidedly stringy way. The prescription is the following:
\begin{equation}
\left<\prod\limits_{i=1}^n  \phi^\dagger_{a_i}\phi(p_i)^{\phantom{\dagger}}_{a_i}\right>=\int\prod\limits_{i=1}^n dt_i\int_0^\infty \frac{dT}{T} \int_{\mathcal{C}} [Dx]\, e^{i \,p_i^\mu x_\mu(t_i) }  e^{-\int_0^T dt \, \left( \frac{\dot{x}^2}{4} + m^2\right) } \text{Sp.}\int D\theta \, e^{-\sigma S_{\text{Doug.}}}
\end{equation}
we move to momentum space by performing functional Fourier transforms inside the path integration. The "marked points" of the position space amplitude are all associated to a Fourier kernel and then summed over, this is the most natural way to proceed. These Fourier kernels then turn out to look exactly like open string vertex operators, in a portentous fashion.

We can transform the expression for the inserted operators, much like in string theory, in order to proceed with the computation. Let us define a piecewise constant momentum function $q$, defined over the contours at hand, such that
\begin{equation}
	\dot{ q}(t)^\mu= -\sum\limits_{i=1}^n p^\mu_i\delta(t-t_i)\,\, , \,\, p_i^\mu x_\mu(t_i) = -\int dt \,\dot{ q}(t)\cdot x(t) =  \int dt \, q(t)\cdot \dot{ x}(t)
\end{equation}

With some further Fourier manipulations and inserting the Douglas action, we obtain
\begin{align}
&\left<\prod\limits_{i=1}^n\phi^\dagger_{a_i}\phi(p_i)^{\phantom{\dagger}}_{a_i}\right>=\label{momamp}\\
&\int\prod\limits_{i=1}^n dt_i\int_0^\infty \frac{dT}{T} \int_{\mathcal{C}} [Dx]\int\frac{d^dk}{(2\pi)^d}\, e^{i \int dt \, (k+q)\cdot \dot{ x}(t) }  e^{-\int_0^T dt \, \left( k^2+m^2\right)  }\text{Sp.}\int D\theta \, e^{-\sigma S_{\text{Doug.}}}\nonumber
\end{align}
To compute this we first perform the $x$-integral, which, with some integration by parts, is quadratic. We obtain the following:

\begin{align}
&\int\prod\limits_{i=1}^n dt_i\int_0^\infty \frac{dT}{T} \int\frac{d^dk}{(2\pi)^d}\, e^{-\int_0^T dt \,\left(  k^2+m^2\right)  }\times \nonumber\\
&\text{Sp.}\int D\theta \,\exp\left(-\frac{1}{2\pi \sigma}\iint d\tau d\tau'(\dot{k}+\dot{q})\log(\theta(\tau)-\theta(\tau'))(\dot{k}+\dot{q})\right)
\end{align}

Given the definition of $q$ above, the quadratic term in $q$ is independent of parametrisation. After taking the massless limit, this term factors out of the $D\theta$ path integral and produces a term of the form
\begin{equation}
\int\prod\limits_{i=1}^n dt_i \exp\left( -\sum\limits_{j<k}\frac{1}{2\pi\sigma} p_j\cdot p_k \log\left( |t_j-t_k|\right) \right) =\int\prod\limits_{i=1}^n dt_i \prod\limits_{j<k} |t_j-t_k|^{-\frac{1}{2\pi\sigma} p_j\cdot p_k}
\label{eq:kobanielsen}
\end{equation}
which, up to a correct definition of the measure, is the Koba-Nielsen integrand required.

The above integration is not complete though, and in general still has non-trivial computations left to do. Makeenko and Olesen offer two ways to simplify the result, either a very large effective string tension (controlled by the parameter $\sigma$) or a large number of inserted points. In either regime, it is the quadratic piece in $p$ that dominates the integration, whatever remains can be assumed to be $p$-independent and so only contributes to overall normalisation. We will choose the former: this is physically justified, because, from the original lattice perspective, the string tension scales with the QCD coupling, which we assume large. The content of Eq.(\ref{eq:kobanielsen}) above is then the main contribution,  and we obtain the famous Veneziano amplitude for $n=4$ ($B$ is the Euler Beta function):

\begin{align}
\left<\prod\limits_{i=1}^n  \phi^\dagger_{a_i}\phi(p_i)^{\phantom{\dagger}}_{a_i}\right> &\propto \int\limits_{t_1<\dots<t_n} \prod\limits_{i=1}^n dt_i \prod\limits_{k<l} |t_k-t_l|^{\frac{1}{2\pi\sigma} p_i \cdot p_j}\\
&= B\left( -\frac{1}{2\pi\sigma} p_1\cdot p_2,-\frac{1}{2\pi\sigma} p_1\cdot p_3 \right) \,\,\, (\text{if }\,\,n=4)
\end{align}

At this point we note that there is a fundamental conceptual difference between this computation and the flat-space open string amplitude it claims to reproduce. Thinking geometrically, Makeenko and Olesen's derivation sums over all shapes and sizes of a certain closed curve, but systematically choose a surface of minimal area whose boundary is that curve. In string theory the path integration is over all coordinates, in the bulk and the boundary of the worldsheet, so therefore necessarily over \textit{all} surfaces for every contour. However, we perform this identification with the Veneziano amplitude only at large values of the string tension, which is precisely the condition to enforce in string theory in order for minimal areas to dominate in the integration. 

It is not clear if the result still holds when we include non-minimal surfaces in Makeenko and Olesen's path integral. Certainly, the fact we are inserting operators on the boundary of the worldsheet means that the bulk degrees of freedom somewhat decouple from the overall computation, as was stated in the derivation, but ideally one would like to extend the reach of this computation and try to add quantum deviations of the bulk behaviour. Doing so in the set-up at hand is difficult: we merely substituted an unknown, complicated function (the Wilson loop expecation value) for a simple geometrical ansatz. This gives us no inbuilt source of further terms to progressively include effects from non-minimal surfaces to the computation, in pure field theory. However, as was noticed recently, in string theory  \cite{Armoni:2016llq}, the holographic prescription gives us a way of doing so neatly, which we will summarize briefly.

\section{The Scattering Amplitude and Holography}

In the previous sections we have shown that, on the lattice, the four point function
 of scalar meson operators is given, in the 't Hooft large-$N_c$ limit, by a sum over Wilson loops that pass via the scattering points $n_1,...,n_4$
\begin{equation}
	\langle \phi(n_1)^\dag \phi(n_1) \cdots \phi(n_4)^\dag \phi(n_4) \rangle_c = \sum_{l}\frac{\kappa^l}{l}\sum_{C_l} \langle W(C_l)\rangle_{YM} \, .
\end{equation}
Moreover, by using a leading order strong coupling assumption, we can write that every Wilson loop is area-behaved:
\begin{equation}
	\langle \phi(n_1)^\dag \phi(n_1) \cdots \phi(n_4)^\dag \phi(n_4) \rangle_c = \sum_{l}\frac{\kappa^l}{l}\sum_{C_l} \exp (-\sigma {\cal A}) \, \label{lattice_scattering}
\end{equation}
with ${\cal A}$ the minimal area of any possible surface whose boundary is $C_l$.

We also showed that in the naive continuum limit the above equation \eqref{lattice_scattering} is given by
\begin{equation}
  \langle \phi(x_1)^\dag \phi(x_1) \cdots \phi(x_4)^\dag \phi(x_4) \rangle_c =\int_0^\infty \frac{dT}{T} \int_{x_i\in\mathcal{C}} [Dx]\,   e^{-\int_0^T dt \, \left( \frac{\dot{x}^2}{4} + m^2\right) } e^{-\sigma \mathcal{A}(\mathcal{C})} \, ,
\end{equation}
where now the scattering points are $x_1, ..., x_4$. Then, we argued (based on the results of Makeenko and Olesen), that after taking the Fourier transform the scattering amplitude is given by the Veneziano amplitude. Thus, by using pure field theory arguments we have shown that the naive continuum limit of the lattice amplitude should be given by the same expression as of the scattering amplitude computed by flat space string theory. Given the geometric nature of the Makeenko-Olesen ansatz, this was perhaps not entirely surprising, as we have mentioned its relation to the string action (see Appendix \ref{appB} for more details). But at least this may seem somewhat accidental. We would like to show that it is not.

In this section we wish to make the same statement about the equivalence between lattice at strong coupling and string theory, based on arguments from the gauge/gravity correspondence. For simplicity we will assume massless quarks in the application of this duality.

Based on the prescription of calculating expectation values of Wilson loop operators \cite{Rey:1998ik,Maldacena:1998im}, it was argued in \cite{Armoni:2015nja} that the sum over Wilson loops is given by a sum over string worldsheets with the topology of a disk, where the boundary of the disk terminates on the boundary of the AdS space
\begin{align}
 \int_0^\infty \frac{dT}{T}  \int_{x_i\in\mathcal{C}} [Dx]\,   e^{-\int_0^T dt \,  \frac{\dot{x}^2}{4} } \left<\mathcal{W}_T[A]\right>   = \int [DX] [Dg] \exp\left( -{\sigma}{\int d^2\tau\, G_{MN} \partial_\alpha X^M \partial _\beta X^N g^{\alpha \beta}} \right ) \, ,
\end{align}
with $G_{MN}$ the 10d metric of the superstring.

We will argue that when Wilson loops admit an area law the above relation becomes
\begin{align}
 \int_0^\infty \frac{dT}{T}  \int_{x_i\in\mathcal{C}} [Dx]\,   e^{-\int_0^T dt \,  \frac{\dot{x}^2}{4} } \exp (-\sigma {\cal A})  = \int [DX] [Dg] \exp\left( -{\sigma}{\int d^2\tau\, \partial_\alpha X_\mu \partial _\beta X^\mu g^{\alpha \beta}} \right ) \, \label{flat-space}
\end{align}
with $\mu=0,1,2,3$, namely 4d flat space coordinates.

In order to justify the conjectured relation \eqref{flat-space}, let us recall the general features of Wilson loop operators and minimal surfaces in string theory, for the special case of confining gauge theories.

We consider that the Wilson loop is drawn on the boundary of an Anti-de Sitter-like space. To compute its expectation value in field theory, at strong coupling, the holographic prescription imposes that we should integrate over all string worldsheets that hang from this shape drawn on the boundary, into the bulk of space. The minimal surface is viewed \cite{Drukker:1999zq} as the saddle point of this integration over string worldsheets, and corresponds to the classical string theory result of the computation. As a strong-weak duality, we expect that the leading classical gravity solution computes the answer for the leading strong coupling behaviour in field theory. This result holds as long as the total space that we choose has certain features associated with confinement physics.

In order for a string background to exhibit confinement-like behaviours, a set of conditions are known, strongly constraining the shape of the space \cite{Kinar:1998vq}. Notably, they usually admit an infra-red cut-off, in the form of an ``end of space'' or a black hole horizon. When the loop is large the string worldsheet accumulates on the IR cut-off and the result is an area law.

Let us consider a confining background and take a limit where the IR cut-off is taken towards the boundary (usually seen as the UV cut-off). In this limit, all string worldsheet, however small, admit an area law as a solution of their classical equations of motion, as the distance from the boundary to the end-of-space surface is negligible. 
\begin{align}
  \int [Dg] \exp\left( -{\sigma}{\int_{\mathcal{S}} d^2\tau\, \partial_\alpha X_\mu \partial _\beta X^\mu g^{\alpha \beta}} \right ) = \exp(-\sigma {\cal A(\mathcal{S})}) \,.
\end{align}
Now, the string tension of this purported dual model should be influenced by the value of the gauge coupling of the field theory we are interested in, and the strong coupling regime corresponds to high string tension. This means we place ourselves in the classical regime of string theory, and so we do not include fluctuations around this set of minimal surfaces. This is exactly the physics of the lattice strong coupling expansion at large-$N_c$: all Wilson loops, however small, contribute proportionally to the area of the minimal surface it encloses. Geometrically the two computations perform the same operations.

Let us demonstrate the above procedure by using a specific example: Witten's model for large-$N_c$ Yang-Mills theory. The metric is given by
\begin{align}
 ds^2&=g(U)\left(d X^2+\text{d$\tau $}^2 f(U)\right)+\frac{1}{g(U)}\left( \frac{d U^2 }{f(U)}+ U^2d\Omega ^2 _4\right) \nonumber\\
 &\text{where }g(U)=\left(\frac{U}{R}\right)^{3/2}\textbf{, }f(U)=1-\frac{U_{\text{KK}}^3}{U^3}
\label{action}.
 \end{align}
Let us take the limit $U_{KK}\rightarrow \infty , R\rightarrow \infty$ with $U_{KK}/R$ fixed. The dynamics in the vicinity of $U \sim U_{KK}$ is that of flat space. The scattering amplitude of mesons is given in momentum space by the Veneziano amplitude. This kind of process can be performed for most dual spaces of relevance, this is known by the analysis performed in \cite{Kinar:1998vq} cited previously, due to the conditions expressed for confining dual string backgrounds. A detailed proof of this remark was offered in \cite{Armoni:2016llq}.

Thus, the picture is as follows. In the lattice side the amplitude is governed by a area behaved Wilson loops. There is no trace of asymptotic freedom or a non-trivial RG flow between the Gaussian UV fixed point and confinement in the IR. The physics of the string side is the same: by taking the IR cut-off to the UV, we focus on physics in the vicinity of the IR. The expectation value of all Wilson loops, however small, can be calculated by classical flat space string theory - every loop contributes proportionally to the area of the minimal surface that it encloses.

The equivalence between the lattice computation and its stringy counterpart is restricted to large-$N_c$, strong coupling/high tension and massless case. It is demonstrated in coordinate space, though in the string side it is more natural to calculate amplitudes in momentum space.

\section{Conclusions}

In this paper we discussed scattering amplitudes of mesons in lattice gauge theory and in flat space string theory. We compared the calculation of an $n$-point function of meson operators and demonstrated that the naive continuum limit of the expression obtained from the lattice coincides with the expression obtained from string theory.

The calculation is restricted to the 't Hooft / planar limit of massless scalar chromodynamics in the strong coupling regime. It tells us that, in this limit, confining gauge theories on the lattice show string-like behaviour at a deep level. This is, by itself, not entirely surprising, since the physics of the lattice strong coupling limit involve sums over rigid (minimal) surfaces. We expect that the analysis could be extended to QCD with fermions, rather than scalars: the leading features of the Veneziano amplitude always appear, in string theory, whatever the spin of the states we choose to scatter.

It would be interesting to ask whether the coincidence of the amplitudes holds beyond the massless limit, beyond the 't Hooft limit and beyond the strong coupling / high string tension limit. In particular a comparison of worldsheet fluctuations (the L\"uscher term). Also, on the string side, some progress was recently made in including non-minimal areas to the computation \cite{Armoni:2016llq}.

It is also interesting to know if there is a stringy worldsheet formulation of the lattice theory. Such a framework can lead to a better understanding of both QCD and its string dual.

We postpone these questions for future work.

\vskip 1cm

{\bf Acknowledgements} We would like to thank Agostino Patella, Jonas Gleeasen and Michele Caselle for stimulating discussions. We thank Biagio Lucini for useful comments on the manuscript. The work of DV is partly supported by the STFC grant ST/L000369/1.

\clearpage
\begin{appendices}
	\section{Gauge fixing reparametrisation invariance}
\label{appA}
In textbook derivations e.g. in \cite{Itzykson:1989sx} it was noted that the worldline-type expressions one gets out of taking the continuum limit of certain distributions of random walks do not seem to be reparametrisation invariant. We certainly expect them to have that property: this formalism introduces an arbitrary parameter, the worldline parameter, the details of which should not lead to any physical differences. Summing naively over all functions $x(t)$ of an interval into spacetime overcounts geometrical objects, because these functions can describe the same geometrical curve in two different parametrisations. 

These comments made about the lattice continuum limit miss an important assumption in their derivation: in all of the above, the parameter $T$ is assumed to be the total proper length of a particle's trajectory, and implicitly also the range of the worldline parameter $t$. The lattice continuum limit integral is therefore in a gauge-fixed form, where a specific coordinate system for the worldline is chosen, and it is proper-length parametrisation. Then, every factor of the worldline metric or einbein vanishes: by definition, the proper-length coordinate is exactly the one for which the metric is unity. Their appearance can be restored in the expressions above, in the right places, in order to satisfy reparametrisation invariance.

It is relevant here to re-derive the correct definition of these sums over all geometric paths, showing explicit reparametrisation invariance. This proof was provided by Polyakov \cite{Polyakov:1987ez}, we will summarise it here.

The action of a free massive particle is the following
\begin{equation}
	S=m\int dt \sqrt{\dot{x}^2(t)}= m T
\end{equation}
where $T$ is the proper length of the path we consider. This action, when minimized by varying $x^\mu(t)$, produces the geodesic equation, as is well-known. The function $x^\mu$ maps an interval, which can be taken \textit{a priori} to be $[0,1]$, into a certain, potentially curved spacetime. Such functions are called embeddings, and we usually specify their boundary conditions: either we fix two end points in spacetime that all possible embeddings have to start and end on, or we impose that the embedding is periodic over the interval, producing closed curves. We will focus on the latter case.

We are interested in using this action as the weight of a statistical ensemble, i.e. to use this action in a path integral, defining a partition function. It would allow us to compute sums and averages of quantities defined over all possible paths through space, weighted by the inverse exponential of their length. Because of reparametrisation invariance, however, if we want to use this action as a weight for path integration, we have to define the integration measure properly. Any change of variable $\tilde{t}=f(t)$ leaves the action invariant for any monotonically increasing function $f$ that does not alter the endpoints. 

 The statistical ensemble we want to be summing over is the set of all geometrically distinct curves, not the set of all embeddings $x(t)$. In other words, we should not sum over all the ways to reparametrise every curve that can be traced in spacetime, as this space is physically meaningless and infinite in functional volume. Since the action above is invariant under reparametrisations, there are ways of removing this unwanted sum.

In practice, this means that the integration measure needs to be defined in the following way. The set of all possible embeddings $x(t)$ can be split up into the product of the set of all geometrically distinct curves, and the set of functions $f$ that reparametrise each of these curves. This set of all geometrically distinct curves is then coset, or set of all equivalence classes, of embeddings under identification by reparametrisation. We want to compute the measure of this coset, $\frac{Dx}{Df}$. Only then can we write a partition function using the action above. We can then write, symbolically
\begin{equation}
	\mathcal{Z}=\int \frac{Dx}{Df}\, e^{-S[x]}
\end{equation}
and perform the computations we desire with this object.

Dividing up the measure as done above is a well understood procedure in quantum field theory, much as the procedure to turn its non-quadratic form into something more prone to computations. Like in string theory, where the same problem exists, Polyakov proposes a solution: to make it quadratic by introducing an auxiliary, arbitrary metric for the curve, whose values we path integrate over. Its value is then forced to be the actual trajectory's metric through its own equation of motion. This can be done this way
\begin{align}
\int \frac{Dx}{Df} \exp\left(-m\int dt \sqrt{\dot{x}^2(t)} \right) &= \int Dx\int \frac{ Dh}{Df} \exp\left(\int dt \sqrt{h(t)} m^2 + \frac{1}{4\sqrt{h(t)}}\dot{x}^2(t) \right) 
\end{align}
That is, the equation of motion of $h$ eliminates this extra variable and reproduces the previous result. 
We can reorder the above as
\begin{align}
\int \frac{Dx}{Df} \exp\left(-m\int dt \sqrt{\dot{x}^2(t)} \right) &= \int \frac{ Dh}{Df} \exp\left(\int dt \sqrt{h(t)} m^2\right) \int Dx \exp \left(\frac{1}{4\sqrt{h(t)}}\dot{x}^2(t) \right)~. 
\end{align}
so that we can treat $\frac{ Dh}{Df}$ first and foremost.

Since any reparametrization will leave the proper length invariant, the measure $Dh$ can be split up like this:
\begin{equation}
	Dh=dT\times Df\times \text{(a Jacobian)}
\end{equation}
where $T$ is again the proper length of the given path. We expect an integral of the form $\int_0^\infty dT$ to appear in our expressions.

 The Jacobian introduced above is an ill-defined operator determinant, it is divergent. As we are not interested in performing an explicit discretisation of the measure $Dx$ we will not do so, and so we can discard the contributions of this Jacobian altogether. The operators we insert are not affected by the functional whose determinant we are computing. 
 
 Finally the term $Df$ is the unwanted integration measure: the one of the space of all reparametrisations. We want to remove it, but, since we want to focus on closed curves and periodic functions,  we have some additional subtleties to capture. One, affecting the integral $\int Dx$, is a zero mode of the action ($x^\mu(t)=c^\mu$ is a trivial solution to the equations of motion, due to translation invariance of spacetime), the other is shift symmetry of the parameter ($t\rightarrow t+a$ leaves the action invariant).
 
 The zero mode, related to translations of the field $x$, is useful, much like in string theory: when we insert operators of definite momentum, integration over the zero mode guarantees that external momentum is conserved. We will not take pains to remove it, it derives from properties of spacetime, which we are not addressing here. On the contrary, the shift symmetry is part of the group of diffeomorphisms on the worldline, and so needs to be removed along with the others. We plan on fixing gauge invariance by computing $\frac{Dh}{Df}$, but we must be careful- the translation above does not change the metric, so the above approach will fail to capture it. Therefore, we remove it by hand before anything else happens, by making use of the relation
 \begin{equation}
 1=\int_{0}^{T} da\,\,\int_0^T \frac{dt}{T} \delta(f(t)-a)
 \end{equation}
 which we insert in the path integral, and drop the integration over $a$. This then allows us to pick any value of $a$ we want, e.g. $0$. Thus, the invariance under shift symmetry is fixed by adding
 \begin{equation}
 \frac{1}{T} \int_0^T dt\, \delta(f(t))
 \end{equation}
 to our path integration. Performing the integral symbolically removes this shift symmetry from the space. We are left with this factor of $T^{-1}$ which effectively allows us to compensate for the overcounting of every curve due to the fact we are not explicitly choosing a starting point.

We can now then drop the $Df$ integration, which allows us to pick any $f$ we please, equivalently it means we can choose a parametrisation, which we pick to be proper time for simplicity and elegance: this sets $h$ to $1$ and the bounds of integration of the worldline parameter to $[0,T]$. The final result is 

\begin{equation}
	\int_0^\infty \frac{dT}{T}\int_{x(0)=x(T)} [Dx]   \exp\left(-\int_0^T dt \,\left( \frac{\dot{x}^2}{4} + m^2\right)\right) 
\end{equation} 

This defines the statistical ensemble of random closed curves that we are integrating over (up to a zero mode contribution which is geometric in nature and so should remain). We use the symbol $[Dx]$ to remind ourselves that this measure has been gauge-fixed, that we have chosen proper length parametrisation in all the expressions we encounter, as is reflected by the appearance of the invariant length $T$ in bounds of integration.

\section{Free strings with fixed boundaries}
\label{appB}
The Douglas functional described  in Eq.(\ref{douglas}) is somewhat difficult to rigorously, mathematically justify, but some sense can be made of it through free string path integration. We impose a fixed spacetime boundary for a theory of open strings described by the Polyakov action, and integrate out the degrees of freedom describing the inside of the worldsheet: the result is the Douglas functional. The proof is roughly presented by Fradkin and Tseytlin \cite{Fradkin:1985qd}, glossing over some finer points. Let us parametrise the worldsheet by upper half-plane variables: $\sigma$ runs from $0$ to $\infty$ and $\tau$ runs from $-\infty$ to $+\infty$.

We then split up the string embedding variable this way:
\begin{align}
X(\sigma,\tau)&=X(0,\tau)\delta(\sigma)+X(\sigma,\tau)(1-\delta(\sigma))\\
&\equiv \hat{X}(\tau) + \mathring{X}(\sigma,\tau)
\end{align}
The variable $\hat{X}$ is the embedding of the boundary of the worldsheet and $\mathring{X}$ embeds the interior of the worldsheet. We then introduce a fixed geometric curve in spacetime $\zeta(\tau)$ and propose to compute
\begin{equation}
\int DX\,Dg \, \delta(\hat{X}-\zeta) \exp\left( -\sigma \int d\sigma d\tau \sqrt{g}\,\, \eta_{\mu\nu}g^{\alpha\beta}\partial_\alpha X^\mu \partial_\beta X^\nu\right) 
\end{equation}
To do this, we write the delta function as a functional Fourier kernel
\begin{equation}
\delta(X-\zeta)=\int D\lambda \exp\left(\int d\tau\lambda(\tau)\left(  X(\tau)-\zeta(\tau)\right)\right)
\end{equation}
and split the $X$ coordinate as prescribed, leading to
\begin{align}
\int D\mathring{X}D\hat{X}DgD\lambda\exp\left(\vphantom{\int}\right.&\left.-\sigma \int_{\sigma>0} d\sigma d\tau \sqrt{g}g^{\alpha\beta}\partial_\alpha \mathring{X}\cdot \partial_\beta \mathring{X}\right) \exp\left(-\sigma\int_{-\infty}^{\infty}d\tau \sqrt{g} g^{\tau\tau}\frac{d}{d\tau}\hat{X}\cdot\frac{d}{d\tau}\hat{X}\right.\nonumber \\
&\left.-\lambda(\tau)\left(  X(\tau)-\zeta(\tau)\right)\vphantom{\int}\right)
\end{align}
Note that there is no $\hat{X}\mathring{X}$ cross-term as these variables are never non-zero simultaneously given their definition.

At this point we need to worry about fixing the large amount of gauge symmetry in the path integral. We are interested in inserting operators on the boundary of this worldsheet, so that the $\mathring{X}$ determinant just produces some functional determinants, as long as they are well defined, but the $g$ integral subsists in the boundary action. We would like to simply fix a flat gauge everywhere, as it is often possible when doing tree-level amplitudes in string theory, but we encounter a difficulty- we have, by specifying $\zeta(\tau)$, given not only a geometric object but also a specific parametrisation of the contour as an input. The output should be parametrisation-independent, therefore Polyakov's approach to the problem is to gauge fix everything but the way we parametrise the boundary, i.e. we fix the worldsheet metric to be diagonal and
\begin{equation}
g_{\sigma\sigma}=1\,\,\, , \,\,\, g_{\tau\tau}=\left( \frac{d \theta}{d\tau}\right) ^2
\end{equation}
for some arbitrary reparametrisation $\theta$. This function encapsulates all of the gauge that we cannot fix, because of the fixed boundary condition. Most of the path integration over $g$ is now gauge-fixed, up to a left-over integration over $\theta$.

As previously mentioned the $\mathring{X}$ integral produces a constant determinant, since the operators we are interested in inserting are all on the boundary, this constant drops out by normalisation, so we ignore it. We are left with
\begin{equation}
\int D\hat{X}D\theta D\lambda \exp\left(-\sigma\int_{-\infty}^{\infty}d\tau \frac{d}{d\tau} \hat{X}(\theta(\tau))\cdot\frac{d}{d\tau}\hat{X}(\theta(\tau))
-\lambda(\theta(\tau))\left(  X(\theta(\tau))-\zeta(\theta(\tau))\right)\vphantom{\int}\right)
\end{equation}

We then perform the $\hat{X}$ path integral, which is linear-quadratic, we get
\begin{equation}
\int D\theta D\lambda \exp\left(-\sigma^{-1}\int_{-\infty}^{\infty}d\tau d\tau' \lambda(\theta(\tau))N(\tau,\tau')\lambda(\theta(\tau'))
+\lambda(\theta(\tau))\zeta(\theta(\tau))\vphantom{\int}\right)
\end{equation}
where $N(\tau,\tau')$ is proportional to the string Neumann function (the Green's function for the string Laplace operator), restricted to the boundary of the worldsheet. Then, integrating out again gives 

\begin{equation}
\int D\theta \exp\left(-\sigma\int_{-\infty}^{\infty}d\tau d\tau' \zeta(\theta(\tau))N^{-1}(\tau,\tau')\zeta(\theta(\tau'))\right)
\end{equation}

For our upper half plane representation of the worldsheet, we have
\begin{equation}
N(\tau,\tau')=\log(|\tau-\tau'|)\,\, , \,\, N^{-1}(\tau,\tau')=\frac{1}{(\tau-\tau')^2}=-\ddot{N}{(\tau,\tau')} 
\end{equation}
where the last property follows from the definition of $N$ as a Neumann function (the inverse power is meant in the functional operator sense, that is, inverse under the convolution product). Using it, integrating by parts and relabelling parametrisations results in the standard Douglas functional above in Eq.(\ref{douglas}).

\end{appendices}
\clearpage
\bibliographystyle{JHEP}

\bibliography{00-main}

\end{document}